
\input epsf
\input phyzzx

\def\ev{{\rm e\kern-.1100em V}}
\def\gev{{\rm G\kern-.1000em \ev}}
\def\tev{{\rm T\kern-.1000em \ev}}
\def\U#1{U\kern-.25em \left( #1 \right) }
\def\su#1{S\U #1 }
\def\up#1{\kern-0.45em\raise0.4em\hbox{$\scriptstyle #1 $}\hskip0.29em}
 \font\bigboldiii=cmbx10 scaled\magstep3
 
 \def\lcal{{\cal L}}
 \def\ocal{{\cal O}}
 \def\ui{U(1)}
 \def\and{{\it\&}}
 \def\half{{1\over2}}
 \def\lowti#1{_{{\rm #1 }}}
 \def\tr{ \hbox{tr}}
 \def\quarter{{1\over4}}
 \def\gesim{\,{\raise-3pt\hbox{$\sim$}}\!\!\!\!\!{\raise2pt\hbox{$>$}}\,}
 \def\lesim{\,{\raise-3pt\hbox{$\sim$}}\!\!\!\!\!{\raise2pt\hbox{$<$}}\,}
 \def\BB{{\bf B}}
 \def\DD{{\bf D}}
 \def\WW{{\bf W}}
 \def\inv#1{{1\over#1}}
 \def\vev{vacuum expectation value}
 \def\ibid{{\it ibid.}\ }

\def\theabstract{We investigate the use of effective Lagrangians to describe
the effects on high-precision observables of physics beyond the Standard Model.
Using the anomalous magnetic moment of the muon as an example, we detail the
use of effective vertices in loop calculations. We then provide estimates of
the sensitivity of new experiments measuring the muon's $ g - 2 $ to the scale
of physics underlying the Standard Model.}

\def\Icol{
\PHYSREV
\nopubblock{
\singlespace
\rightline{$\caps UCRHEP-T98$}
\rightline{$\caps UM-TH-92-17$}
\rightline{$\caps NSF-ITP-92-122i$}}
\rightline{hep-ph/9304206}
{\titlepage
\title{ {\bigboldiii
Effective-Lagrangian approach to precision
measurements: the anomalous magnetic moment of the muon}}
\smallskip
\titlestyle{{\twelvecp C. Arzt \hbox{and} M.B. Einhorn }}
{\twelvepoint\it\centerline{ Institute for Theoretical Physics, University of
California} \baselineskip=12pt \centerline{ Santa Barbara, CA 93106, USA}}
\baselineskip=12pt
\smallskip
\centerline{and}
\smallskip
{\twelvepoint\it\centerline{ Randall Laboratory of Physics, University of
Michigan} \baselineskip=12pt \centerline{ Ann Arbor, MI 48109-1120,
USA\foot{{\rm Present address}} }}
\baselineskip=12pt
\medskip
\centerline{and}
\medskip
\titlestyle{{\twelvecp J. Wudka }}
{\address{{\it University of California at Riverside\break
 Department of Physics\break
 Riverside{\rm,} California 92521{\rm--}0413{\rm,}
 U{\rm.}S{\rm.}A{\rm.} \break
 WUDKA{\rm @}UCRPHYS}}}
\singlespace
\abstract
\theabstract
\endpage}
\doublespace}

\def\amu{a_\mu}
\def\anew{\delta a_\mu}
\def\anewd{ \anew \up{{\rm direct}} }
\def\anewi{ \anew \up 1 }
\def\anewii{ \anew \up 2 }
\def\lef{\lcal \lowti{ eff } }
\def\sw{s \lowti w }
\def\cw{c \lowti w }

\Icol

\chapter{{\caps Introduction}}

One of the time-honored methods of probing physics beyond the known realm is to
perform highly accurate measurements of precisely predicted observables. Rather
than observing new particles directly we infer their existence from the effects
they have on known particles; interactions unexplainable within the framework
of the accepted theory (which in our case is the Standard Model) imply new
physics. This type of experiment, by its very nature, cannot unambiguously
discriminate among the manifold possible models which extend the Standard Model
to higher energies; there may be several competing models which affect measured
quantities in the same way, but no observable to which only one of the models
contributes.

Faced with this situation it becomes questionable to try to understand the low
energy effects of new physics on a model-by-model basis. The best
approach is to follow a characterization which is
sufficiently general to encompass all types of high-energy physics.
Such a characterization is readily available in an effective-lagrangian
approach
\REF\ac{
T. Appelquist and J. Carazzone, {\sl Phys. Rev.} {\bf D11} (1975) 2856.
K. Symanzik, {\sl Comm. Math. Phys.} {\bf 34} (1973) 7. }
\REF\gp{
H. Georgi and H. Politzer, {\sl Phys. Rev.} {\bf D14} (1976) 1829.
E. Witten, {\sl Nucl. Phys.} {\bf B104} (1976) 445.
E. Witten, {\sl Nucl. Phys.} {\bf B122} (1977) 109.
J. Collins, F. Wilczek, and A. Zee {\sl Phys. Rev.} {\bf D18} (1978) 242.
S. Weinberg, {\sl Phys. Rev. Lett.} {\bf 43} (1979) 1566.
F. Wilczek and A. Zee, {\sl Phys. Rev. Lett.} {\bf 43} (1979) 1571.
Y. Kazama and Y.P. Yao, {\sl Phys. Rev.} {\bf D21} (1980) 1116.
B. Ovrut and H. Schnitzer, {\sl Phys. Rev.} {\bf D21} (1980) 3369.
S. Weinberg, {\sl Phys. Lett.} {\bf 91B} (1980) 51.}
[\ac,\gp]
which has recently been advocated
\REF\falk{
A.F. Falk \etal, {\sl Nucl. Phys.} {\bf B365} (1991) 523.}
\REF\ewtwo{
M.B. Einhorn and J. Wudka, in: Proceedings of the
{\sl Workshop on Electroweak Symmetry Breaking} Hiroshima, Nov. 12--15 1991.
H. Georgi, {\sl Nucl. Phys.} {\bf B361} (1991) 339;
\ibid\ {\bf B363} (1991) 301.
J. Wudka, lecture presented at the {\sl Topical conference on Precise
Electroweak Measurements}, Santa Barbara, CA 21--23 Feb. 1991.
A. DeR\'ujula {\it et al.}, CERN preprint CERN-TH.6272/91}
\REF\holdt{
B. Holdom and J. Terning, {\sl Phys. Lett.} {\bf B247} (1990) 88.
B. Holdom, {\sl Phys. Lett.} {\bf B258} (1991) 156;
\ibid\ {\bf B259} (1991) 329.}
[\falk,\ewtwo,\holdt]
as a model- and process-independent parameterization of deviations from the
Standard Model.

The effective-lagrangian method should be contrasted with an approach in which
a given model (or set of models) extending the Standard Model is chosen, and
the effects on low energy observables are calculated.  The specific model
approach determines all corrections to the Standard Model in terms of a few
couplings and masses. The effective-lagrangian approach parametrizes its
predictions in terms of the coefficients of effective operators.  Therefore we
are faced with a trade off: the requirement of model independence increases the
number of unknown parameters whose order of magnitude can at best be estimated.
The results of the effective lagrangian approach are very useful in
determining, in a model-independent manner, the sensitivity of a given
experiment to new  physics, and can be used to isolate those observables most
sensitive to possible new interactions.

The basic idea of the method is that processes below some energy $\Lambda$ can
be described by effective operators consisting of fields with masses below
$\Lambda$. From these operators we hope to infer the existence of particles
with masses above $\Lambda$. Thus, so long as we are below all new particle
thresholds, any type of new physics can be parametrized by a series of
effective operators involving Standard Model particles. It is important to note
that any given model will produce operators which respect the (exact)
symmetries of the Standard Model and that the best we can hope for from high
precision measurements are statements regarding the coefficients of these
operators.

The underlying physics is described by a high-energy
lagrangian, out of which all excitations with masses above $
\sim 100\ \gev $ (which we label ``heavy'') are integrated out; what remains
will be the Standard Model, plus an infinite series of effective operators.
These must be gauge-invariant[\ewtwo] \REF\ew{M.B. Einhorn and J. Wudka, {\sl
Phys. Rev.} {\bf D39} (1989) 2758.} \foot{{This is not a trivial result: its
derivation requires the introduction of a gauge fixing technique which produces
a manifestly gauge invariant effective action [\ew].}} and describe the
low-energy remnants of the full high-energy theory. We will denote by $\Lambda$
the (large) energy scale at which the new physics first directly manifests
itself.

There are two possible types of high-energy physics to consider, that which
decouples from low-energy physics and that which does not. In the decoupling
scenario, the masses of heavy degrees of freedom are large because a
dimensionful parameter respecting the symmetries of the theory is large. In
this case the decoupling theorem [\ac] tells us that at energies $E \ll
\Lambda$ corrections to the low-energy theory are suppressed by powers of
${1\over\Lambda}$ (times possible powers of $\ln \Lambda$). In the
non-decoupling case, the masses of the heavy degrees of freedom in the theory
are large because some dimensionless coupling constant is large. An example of
this is a heavy fermion which gets its mass from spontaneous symmetry breaking
and becomes heavy due to a relatively large Yukawa coupling.\foot{The coupling
may still be within the perturbative regime.}  In this case the contributions
due to physics above $\Lambda$ need not be suppressed by powers of
${1\over\Lambda}$; the corrections to Standard Model processes are given by a
chiral expansion in powers of $ p /\Lambda  $, where $p$ is a typical
momentum for the process at hand\REF\georgi{H. Georgi, {\sl Weak Interactions
and Modern Particle Theory}, Benjamin / Cummings Publishing Co., Menlo Park,
CA, USA (1984).} [\georgi].

The application of effective-lagrangian techniques to the case of high
precision measurements contains a new complication, for the effective vertices
can appear within loops and thereby produce new divergences.  But the
effective lagrangian is completely renormalizable -  power counting arguments
similar to those involved in proofs of renormalizability show that all
divergences multiply local operators.  Since the effective lagrangian includes
all operators respecting the symmetries of the theory, such divergences simply
renormalize the bare coupling constants.\foot{Note that this does not require
that the underlying theory be renormalizable.} Their only effect (associated
with the logarithmic divergences) is to determine the renormalization-group
running of the effective couplings.  This is a well known fact and has been
applied in the context of the strong interactions \REF\gl{J. Gasser and H.
Leutwyler, {\sl Ann. Phys.} {\bf 158} (1984) 142; {\sl Nucl. Phys.} {\bf B250}
(1985) 465; \ibid {\bf B250} (1985) 517; \ibid {\bf B250} (1985) 539.  S.
Weinberg, {\sl Physica} {\bf A96} (1979) 327.}[\gl].

We will develop and apply the techniques of effective lagrangians using the
anomalous magnetic moment of the muon as an example. This is an especially
interesting observable because the Brookhaven experiment AGS 821
\REF\hughes{V.W. Hughes, AIP conference proceedings no. 187, (1989) 326.  M.
May, AIP conference proceedings no. 176, (1988) 1168.}[\hughes] is expected to
achieve a precision greater than that required to observe the Standard Model
contributions. The best present measurement of $\amu$ comes from a series of
experiments at CERN \REF\bailey{J. Bailey \etal., {\sl Nucl. Phys.} {\bf B150}
(1979) 1. E.R. Cohen, B.N. Taylor, {\sl Rev. Mod. Phys.} {\bf59} (1987) 1121.}
[\bailey]: $$ a_{CERN} = 11 \ 659 \ 230 (84) \times 10^{-10}. \eqn\eq $$ The
calculated effects on $\amu$ of weak interactions are \REF\bgl {W.A. Bardeen,
R. Gastmans, and B. Lautrup, {\sl Nucl. Phys.} {\bf B46} (1972) 319. R. Jackiw
and S. Weinberg {\sl Phys. Rev.} {\bf D5} (1972) 157. I. Bars and M. Yoshimura
{\sl Phys. Rev.} {\bf D6} (1972) 374. J. Calmet, \etal, {\sl Rev. Mod. Phys.},
{\bf49} (1977) 21.} [\bgl]: $$ a_{weak} \simeq 19.5 \times 10^{-10},
\eqn\eq $$ too small to be seen in the CERN experiment, but well within the
projected accuracy of the AGS experiment, which should measure $\amu$ with an
accuracy of $ 4 \times 10^{-10}$.

Several authors \REF\km{T. Kinoshita and W.J. Marciano, Cornell
University report CLNS 90-983, to appear in {\sl Quantum Electrodynamics}, T.
Kinoshita, ed. (World Scientific, Singapore, 1990). P. M\'ery, S.E. Moubarik,
M. Perrottet, F.M.Renard, {\sl Z. Phys} {\bf C46} (1990) 229.
A.I. Studenikin
and I.M. Ternov, {\sl Phys. Lett.} {\bf B234} (1990) 367. M. Suzuki, {\sl Phys.
Lett.} {\bf B153} (1985) 289. A. Grau, J.A. Grifols, {\sl Phys. Lett.} {\bf
B154} (1985) 289. J.C. Wallet, {\sl Phys. Rev.} {\bf D32} (1985) 813.}
[\km] have considered the effects on $\amu$ of gauge-boson anomalous
magnetic dipole and electric quadrupole moments. We will re-examine this
problem using the effective-lagrangian formalism to discuss the effects of
high-energy physics on these constants and will examine the previous results
from this point of view.

Regardless of whether or not the heavy excitations decouple, the contributions
to $ \amu $ produced by the underlying interactions can be classified in three
types: $$ \eqalign{ \anewi &: \hbox{ produced \ by \ loops \ containing \ the
\ effective \ operators. } \cr \anewii &: \hbox{ produced \ by \ the \
modification \ of \ the \ gauge \ boson \ eigenstates.} \cr \anewd &:
\hbox{produced \ by \ effective \ operators \ of \ the \ type} \ \bar \mu
\sigma_{\mu \nu } \mu F^{ \mu \nu }. \cr} \eqn\class$$ The contribution $
\anewii $ is due to the fact that the new interactions modify the quadratic
part of the lagrangian, so that a re-diagonalization of these terms is
required. We will consistently use this classification below.

\chapter{{\caps Decoupling case}}

As mentioned above, in the decoupling scenario the low-energy effective
lagrangian can be written as a power series in $ 1/ \Lambda $, $$ \lef =
\sum_{n=0}^{\infty} {1\over{{\Lambda}^n}}\thinspace \alpha_{\ocal}\thinspace
\ocal^{(n+4)} \eqn\eq $$ where the operators $\ocal^{(n+4)}$ have dimension
$[\hbox{mass}]^{(n+4)}$, are $ \su2_L \times \ui_Y $ gauge invariant, and
contain only Standard Model fields. The constants $\alpha_{\ocal}$ (which must
be renormalized) determine the strength of the
contribution of $\ocal$. We shall assume that in this expansion $\ocal^{(4)}$
is equal to the Standard Model, which then takes the status of an effective
lagrangian valid for energies much less than $\Lambda$; this is true only in
the decoupling case. As the structure of the interactions underlying the
Standard Model are presently unknown, the coefficients $ \alpha_\ocal $ cannot
be evaluated; nonetheless, their order of magnitude can be estimated [\georgi].

The determination of the contribution to $ \amu $ from dimension-six operators
is simplified by the constraint that these operators must be gauge invariant.
Additionally, we may use the classical equations of motion to remove some
redundant operators
\REF\arzt{C. Arzt, University of Michigan preprint UM-TH-92-28}
\ [\arzt].
Buchm\"uller and Wyler
\REF\bw{W. Buchm\"uller and D. Wyler, {\sl Nucl. Phys.}
{\bf B268} (1986) 621; {\sl Phys. Lett.} {\bf B197} (1987) 379. See also:
C.N. Leung, \etal, {\sl Z. Phys.} {\bf C31} (1986) 433. C.J.C. Burgess and H.J.
Schnitzer, {\sl Nucl. Phys.} {\bf B228} (1983) 464.}
\ [\bw] have compiled a list of all possible gauge-invariant terms (assuming
lepton and baryon number conservation) in an effective lagrangian to order $
1/ \Lambda^2 $. They find that there are no dimension five operators and 81
operators of dimension six (for a single fermion family).

In determining $ \anew \up{ 1 , 2 } $ (see \class) we will not treat all
dimension six operators, but will instead focus on those operators which give
rise to anomalous three-gauge-boson or two-gauge-boson-Higgs couplings, and
thereby to anomalous gauge-boson dipole and electric quadrupole moments.
There are four such operators in [\bw]
$$ \eqalign{
\ocal_W &= - \epsilon_{ I J K} { W_\mu }^{ I \nu } { W_\nu }^{ J \lambda }
                   { W_\lambda }^{ K \mu }, \cr
\ocal_{ W B } &= ( \phi^\dagger \tau^I \phi ) { W_{ \mu \nu } }^I
                   B^{ \mu \nu}, \cr
\ocal_{ \phi W } &= \half ( \phi^\dagger \phi ) W_{ \mu \nu }^I
                   W^{ I \mu \nu }, \cr
\ocal_{ \phi B } &= \half ( \phi^\dagger \phi ) B_{ \mu \nu }
                  B^{ \mu \nu } . \cr }
\eqn\effop
$$
$\ocal_{W}$ and $\ocal_{WB}$ contribute to figure 1 (a) (there are three
additional graphs where the W bosons are replaced by the corresponding would-be
Goldstone bosons).  $\ocal_{WB}$, $\ocal_{\phi W}$, and $\ocal_{\phi B}$
contribute to figure 1 (b) and (c).  In each figure a heavy dot denotes one of
these effective operators.

\setbox1=\vbox {\hsize=1.5truein
\epsfxsize=\hsize
\epsffile{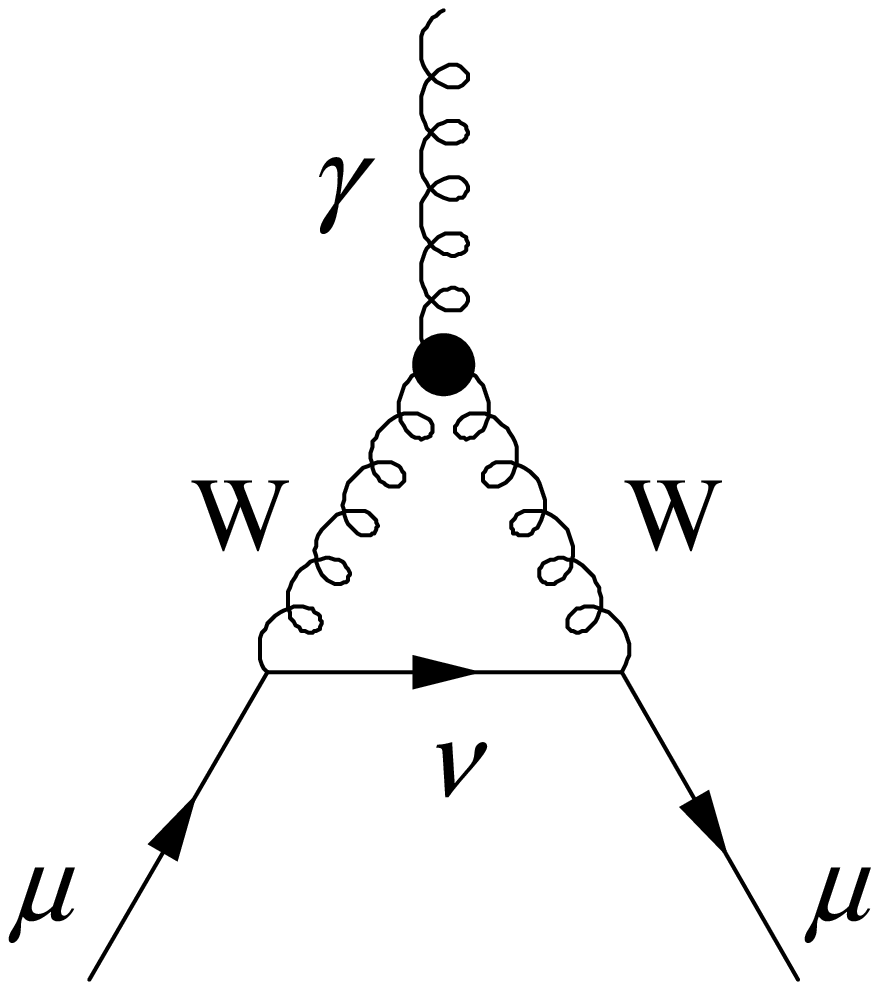}
\singlespace
\centerline {\ninerm (a)}}

\setbox2=\vbox {\hsize=1.5truein
\epsfxsize=\hsize
\epsffile{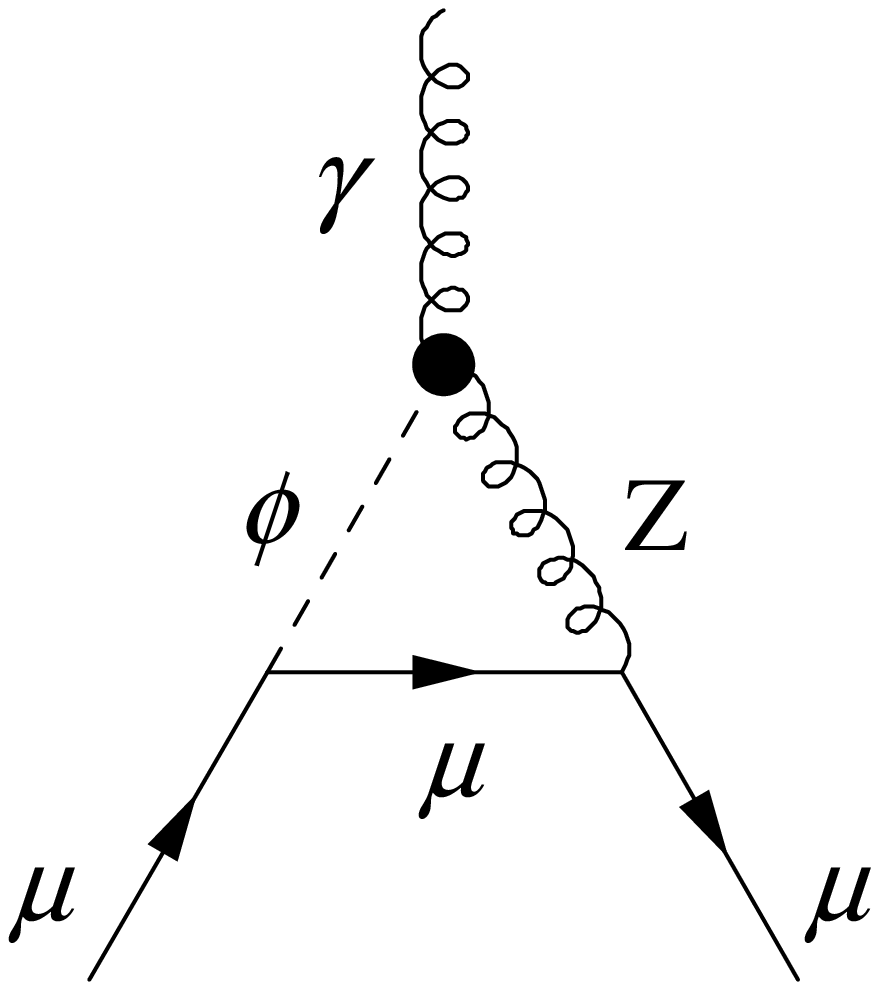}
\singlespace
\centerline {\ninerm (b)}}

\setbox3=\vbox {\hsize=1.5truein
\epsfxsize=\hsize
\epsffile{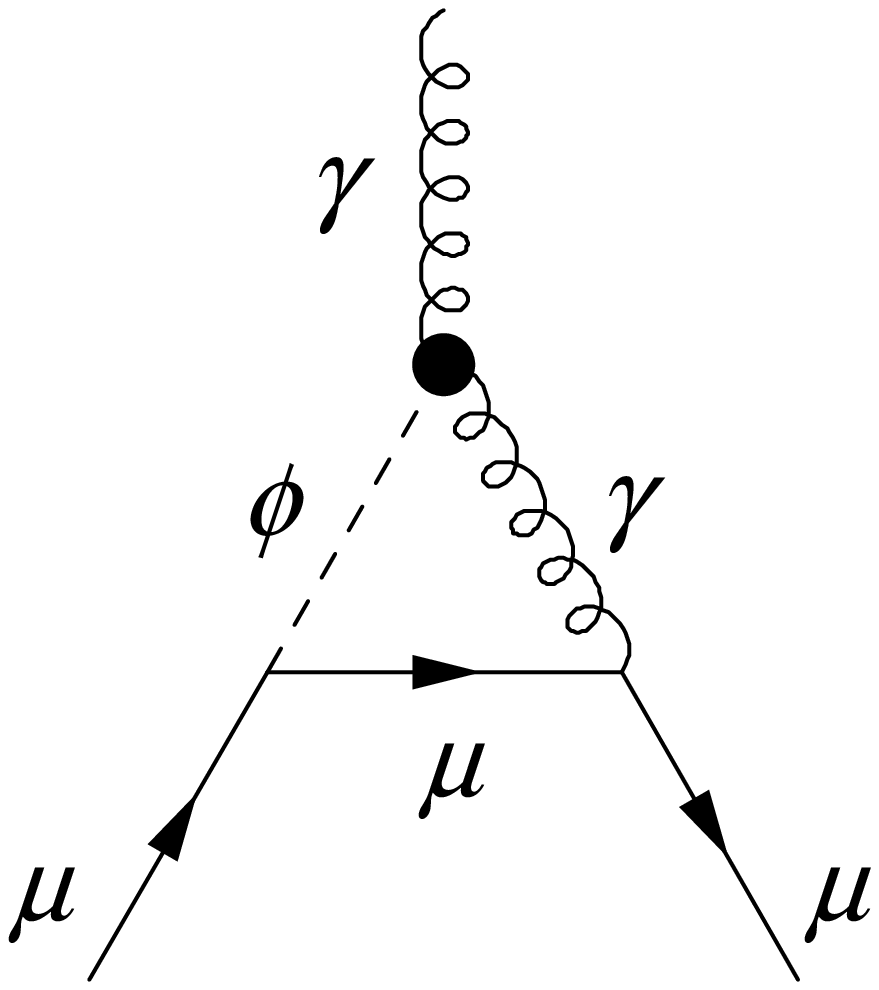}
\singlespace
\centerline {\ninerm (c)}}

\vbox {\line { \hfill \box1 \hfill \box2 \hfill \box3 \hfill}
\centerline {\twelverm Figure~1}}
\vskip.2in

We must also consider the two operators which
can give a direct contribution to $\anew$:
$$
\ocal_{ \mu W }= - ( \bar L \sigma^{\mu \nu } \tau^I \mu_R )
\phi W_{ \mu \nu }^I + h.c.
\quad \hbox{and}
\quad \ocal_{\mu B }= - ( \bar L \sigma^{ \mu \nu } \mu_R )
\phi B_{ \mu \nu } + h.c.;
\eqn\directo
$$
where $ L = { \nu_\mu \choose \mu } $. Our notation is $ W_{ \mu \nu }^a =
\partial_\mu W_\nu^a - \partial_\nu W_\mu^a - g \epsilon_{ a b c } W_\mu^b
W_\nu^c $ for the $ \su 2_L $; $ B_{ \mu \nu } $ for the $ \ui_Y $
field strength, and $ D_\mu = \partial_\mu + ( i / 2 ) g \tau^a W_\mu^a + i g'
Y B_\mu $ ($ Y $ being the hypercharge) for the covariant derivative. $ \phi $
denotes the Standard-Model scalar doublet; its \vev\ is $ \langle \phi \rangle
= { 0 \choose v / \sqrt{2 } } $.

Our effective low-energy bare lagrangian will therefore be
$$
\lef = \lcal \lowti{ S M } + \inv{ \Lambda^2 }
\left( \alpha_W \ocal_W + \alpha_{ W B } \ocal_{ W B } +
  \alpha_{\phi W} \ocal_{\phi W} + \alpha_{ \phi B } \ocal_{ \phi B } +
  \alpha_{ \mu W } \ocal_{ \mu W } + \alpha_{ \mu B } \ocal_{\mu B } \right)
\eqn\ourlef
$$
The above parameterization in terms of the $\alpha_\ocal $ is not standard in
the literature, where the description of $ \lef $ in terms of parameters
frequently called $ \kappa $ and $ \lambda $ are used \REF\hag{ K. Hagiwara,
\etal, {\sl Nucl. Phys.} {\bf B282} (1987) 253.} [\hag]. We will see that $
\alpha_W $ is proportional to $\lambda$, and $ \alpha_{ W B } $ to
$(\kappa-1)$.

When we replace $\phi$ by ${ v\over \sqrt{2}}$, the effects of  $\ocal_{\phi
W}$ and $\ocal_{\phi B}$ on $ \amu $ are not observable. They can be absorbed
into a wave function renormalization of $ B $ and $ W $, together with a
rescaling of the constants $ g $ and $ g' $, and will not be considered
further.  These two operators therefore only contribute to $\amu$ through
figure 1 (b) and (c).

As mentioned previously we do not know the full theory, and so we cannot
compute the six $\alpha$'s, but we can estimate their sizes. It can be shown
\Ref\aew{ C. Arzt, M. Einhorn, and J. Wudka, in preparation} [\aew]
that, because of the SU(2)xU(1) symmetry of the high-energy theory,
the six operators must come from some loop in the full theory with heavy
internal lines. This implies that any $W$ or $B$ must be accompanied by the
corresponding coupling $g$ or $ g'$. Moreover the fact that these operators are
generated by loop graphs implies that the corresponding $ \alpha $ must contain
a factor $ \sim 1/ 16 \pi ^2 $. In this fashion we obtain (for the constants at
scale $\Lambda$)
$$ \eqalign{
\alpha_W \sim { g^3 \over 16 \pi^2 },
\quad \alpha_{ W B } &\sim { g g' \over 16 \pi^2 },
\quad \alpha_{\phi W} \sim { g^2 \over 16 \pi^2 }, \cr
\quad \alpha_{\phi B} \sim { g'^2 \over 16 \pi^2 },
\quad \alpha_{\mu W} &\sim { g \over 16 \pi^2 },
\quad \alpha_{\mu B}  \sim { g' \over 16 \pi^2 } . \cr}
\eqn\estimates
$$
While we can't expect these expressions to be numerically precise, we can use
them as useful order of magnitude estimates. One should, however, be aware of
the possibility that
there may be several such contributions or resonant effects that may
enhance these estimates, perhaps by as much as an
order of magnitude; on the other hand, the presence of small couplings in the
underlying theory can suppress these values below the ones in \estimates.
Moreover, there may even be additional suppression factors of $1/16\pi^2,$
since certain couplings may arise only beyond the one-loop level in the
underlying theory, e.g., the contribution of a fourth generation
to $\alpha_{ \mu W
}$ and $\alpha_{ \mu B }$ sets in at the two-loop level. Thus, the inferences
to be drawn on the sensitivity to the scale $\Lambda$ must be interpreted in
this context and not immediately identified with the threshold for new particle
production.

It might be argued that the coefficients $ \alpha_{ \mu W , B } $ should have a
factor of $ m_\mu / v $ to allow for a natural mass generation for the muon. In
fact, this is not necessarily the case. Supersymmetric models can substitute
for $ m_\mu / v $ a factor of $ m_{\tilde Z} / v $, for example. \foot{This is
not required, though - many supergravity-inspired SUSY models avoid this.} Also
without this factor are models constructed so as to allow a relatively large
magnetic moment for the neutrinos while keeping their mass within experimental
bounds.\REF\vol{ M.B. Voloshin, {\sl Sov. J. Nucl. Phys.} {\bf48} (1988) 512.
G. Ecker \etal, {\sl Phys. Lett.} {\bf B232} (1989) 217. K.S. Babu and R.N.
Mohapatra, {\sl Phys. Rev. Lett.}, {\bf63} (1989) 228. M. Leurer and N. Marcus,
{\sl Phys. Lett.} {\bf B237} (1990) 81.} [\vol].  However, models which to not
contain the $m_\mu / v$ suppression are likely to render the muon mass
unnaturally light.  We shall present results both including and disregarding
this small factor.

When $ \phi $ is replaced by its \vev, then $$ \ocal_{ W B } = - \half { v^2
\over \Lambda^2 } W^3_{ \mu \nu } B^{ \mu \nu } , \eqn\eq $$ which contributes
to the quadratic part of the lagrangian. This necessitates a
re-diagonal\-i\-za\-tion
of the boson fields (see [\bw]) $$ \eqalign{ W^3_\mu =& \sw \left( 1 - \sw \cw
{ v^2 \over \Lambda^2 } \alpha_{ W B } \right) A_\mu + \left( \cw + \sw^3 { v^2
\over \Lambda^2 } \alpha_{ W B } \right) Z_\mu , \cr B_\mu =& \cw \left( 1 -
\sw \cw { v^2 \over \Lambda^2 } \alpha_{ W B } \right) A_\mu - \left( \sw +
\cw^3 { v^2 \over \Lambda^2 } \alpha_{ W B } \right) Z_\mu , \cr } \eqn\fields
$$ where $ \sw $ and $ \cw $ are the sine and cosine of the tree-level weak
mixing angle in the Standard Model. This leads to certain modifications of the
Standard-Model parameters which we reproduce for completeness [\bw] $$ e
\rightarrow e^* = e \left[ 1 - \sw \cw \alpha_{ W B } { v^2 \over \Lambda^2 }
\right] \quad \quad M_Z \rightarrow M_Z^* = M_Z \left[ 1 + \sw \cw \alpha_{ W B
} { v^2 \over \Lambda^2 } \right]. \eqn\eq $$ $ \ocal_{ W B } $ does not affect
$v$, $M_W$ or $ G_F $. We will use $e^*$, $M_Z^*$, and $G_F^*$ as our input
parameters; for example, it is $ M_Z^* $ that is measured to be $ 91.2\ \gev$,
not $ M_Z $.

The Standard Model electroweak one loop contribution to $ a_\mu $ is [\bgl]
$$ a_\mu \up{ SM } = { G_F^* m_\mu^2 \over 6 \sqrt{ 2 } \pi^2 } \left[ 4 \left(
{ M_W^* \over M_Z^* } \right)^4 - 6 \left( { M_W^* \over M_Z^* } \right)^2 + 1
\right] +{ G_F^* m_\mu^2 \over 8 \sqrt{ 2 }\pi^2 }\left({10 \over 3} \right) +
{{e^*}^2 \over 8 \pi^2}. \eqn\amusm $$ The change in the values of $M_Z$ and
the $Z \mu \bar\mu$ couplings results in a change in $a_\mu$; the anomalous
magnetic moment calculated with the shifted fields \fields\ minus the standard
model result \amusm\ is equal to $\anewii$.

The inclusion of effective vertices in loop graphs requires a certain amount of
care: it is assumed from the beginning that all momenta in $ \ourlef $ lie
below $ \Lambda $, and this is violated by the loop momentum in
figure 1.  To deal with this problem, note that when the graphs in
figure 1 are
differentiated once with respect to an external momentum, they become
ultraviolet convergent, and so the momenta entering the heavy loop can be
effectively assumed to be small compared to $\Lambda$; then the use of $
\ourlef $ is justified. Integrating with respect to the above external momentum
produces an undetermined integration constant times a ``direct'' operator
$\ocal_{\mu W}$ or $\ocal_{\mu B}$. We expect that for scales $\mu = \Lambda$
this term will give a contribution of the same order of magnitude as $\anewi$.
In this manner, using the above expressions (and calculating in the Feynman
gauge with dimensional regularization) we obtain from \ourlef\
$$ \eqalign {
\anewd &= { 4 \sqrt{2} M_W m_\mu \over g^2 \sw \Lambda^2 }
       \left( \sw \alpha_{\mu W } - \cw \alpha_{ \mu B } \right) , \cr
\delta a_\mu \up {1a} &= - { 3 g \over 16 \pi^2 } {m_\mu^2 \over \Lambda^2 }
     \alpha_W + \inv{ 8 \pi^2 } { g \over g' } { m_\mu^2 \over \Lambda^2 }
     \alpha_{ W B } \left[ \inv \epsilon - \gamma
     + { 3 \over 2} - \ln { M_W^2\over 4 \pi \mu^2 } \right], \cr
\delta a_\mu \up {1b} &= { 4 \sw^2 - 1 \over 16 \pi^2}
    { m^2 \over \Lambda^2 } \left [\inv\epsilon -
    \gamma + {3 \over 2} +
    {m_h^2 \over M_Z^2 - m_h^2} \ln{ m_h^2 \over 4 \pi \mu^2 } -
    {M_Z^2 \over M_Z^2 - m_h^2} \ln{ M_Z^2 \over 4 \pi \mu^2 } \right ] \cr
 & \ \ \ \ \ \ \ \ \ \ \ \ \ \ \ \ \ \ \ \ \ \ \ \ \ \ \ \ \ \times
   \left( \alpha_{\phi W} - \alpha_{\phi B} +
	        { {\sw^2 - \cw^2} \over \cw\sw } \alpha_{WB} \right), \cr
\delta a_\mu \up {1c} &= - { m^2 \over 4 \pi^2 \Lambda^2 }
    \left [\inv\epsilon -
    \gamma + {3 \over 2} -  \ln{ m_h^2 \over 4 \pi \mu^2 } \right]
    \left(\cw^2 \alpha_{\phi B} +
            \sw^2 \alpha_{\phi W} -2 \sw\cw\alpha_{WB}\right) ,\cr
\anewii &= { m_\mu ^2 \over 6 \pi^2 \Lambda^2 } \alpha_{ W B }
        \sw \cw ( 1 - 4 \sw^2 ) , \cr
}
\eqn\amudec
$$
where $m_h$ is the mass of the higgs, $ \mu $ is the renormalization scale, $
\gamma $ is Euler's constant, and the dimension of space-time is $ 4 - 2
\epsilon $.  We have broken $\anewi$ into three parts, one from each diagram in
figure 1. To this order in $\Lambda$, the starred and unstarred parameters in
\amudec\ are equal. As mentioned above, the divergences are unobservable; the
infinite contributions from the graphs in figure 1 are cancelled by
counterterms of the form of $ \alpha_{ \mu B} $ and $ \alpha_{ \mu W } $.

The complete expression for the new physics contribution to $ \amu $ is given
by the sum of the five contributions in \amudec. We will assume that the
renormalization of the dimension six operators has been carried out using $
\overline{ { \rm MS } } $ \REF\bbdm{W. Bardeen, A. Buras, D. Duke, T. Muta,
{\sl Phys. Rev.} {\bf
D18} (1978) 3998.} [\bbdm] and choose $\mu^2=\Lambda^2$, so that we may use
the  estimates in \estimates. If we renormalize at a different scale, for
example $\mu^2=M_W^2$, then we must run our estimates for $\alpha_{\mu W}$ and
$\alpha_{\mu B}$ from scale $\Lambda$ to scale $M_W$. This running is
determined by the part of $\anewi$ proportional to $\alpha_{WB}$, and the
difference in $\alpha$'s at scale $\Lambda$ and scale $M_W$ will be
proportional to $\ln M_W / \Lambda$ such that $\anewd$ + $\anewi$ is unchanged.
It is only in this sense that the logarithmic divergences in \amudec\ have any
effect. (The couplings $ e , \ g $, and $ g' $ at scale $ \Lambda $ will differ
from those at scale $ M_W $ by terms involving $\ln M_W / \Lambda $, but these
differences are suppressed by higher powers of Standard-Model coupling
constants
and can be ignored to this order.)

We find numerically that
$$ \eqalign {
\anewi + \anewii \simeq
-  1 \times & 10^{-10} { \alpha_W \over \Lambda^2} \cr
+  3 \times & 10^{-9}  \left[ 1 + \inv{6} \ln \Lambda^2 \right]
     {\alpha_{ W B } \over \Lambda^2 }                        \cr
-  4 \times & 10^{-10} \left[ 1 + {1 \over 5} \ln \Lambda^2 \right]
     {\alpha_{ \phi W } \over \Lambda^2 }                        \cr
-  1 \times & 10^{-9}  \left[ 1 + {1 \over 5} \ln \Lambda^2 \right]
     {\alpha_{ \phi B } \over \Lambda^2 }                        \cr }
\eqn\eq
$$
where $ \Lambda $ is to be expressed in
TeV. Then, taking the estimates for the couplings given in \estimates, we
obtain
$$\eqalign {
\left| \anewi + \anewii \right| \simeq
10^{ - 12 } \Bigl|
   & \left[ 4 \pm 1 \pm .9 \pm .3 \right]   \cr
 + & \left[ .8 \pm .2 \pm .2 \right]\ \ln\Lambda^2 \Bigr|
		                        \inv{ \Lambda^2 }  \cr
}
\eqn\loopvalues
$$
where all masses are measured in TeV and the $\pm$ refers to the relative
signs of the various $\alpha$'s. $m_h$ has been set to 150 GeV (the dependence
on $m_h$ is slight).  This equation shows that that the CERN and Brookhaven
experiments are both completely insensitive to $\Lambda$ above the Z mass for
our estimates of $\alpha_W$ and $\alpha_{ W B }$. To reach a bound which is at
all interesting, say $ \Lambda=200\ \gev$, would require the $\alpha$'s to be
about ten times larger than expected. To reach $ \Lambda= 1~\tev$
would require the $\alpha$'s to be about 100 times larger than
expected, an unlikely occurrence in our opinion.

These discouraging results do not apply to $\anewd$ which (assuming no strong
cancellation between the two terms) takes the value $$
\anewd \sim {9 \times 10^{-7 } \over \Lambda^2 } \eqn\treevalue $$ ($ \Lambda $
in TeV) which is about six orders of magnitude larger than the contributions in
\loopvalues. This is because \loopvalues\ contains a factor $ m_\mu / (16 \pi^2
v) \sim 10^{ -6 } $ due to the loop integration and the helicity change of the
muon. Eq. \treevalue\ allows a sensitivity limit of $ \Lambda \lesim 50~\tev$
for the Brookhaven experiment; effects from scales beyond this value will not
be observed. In fact the CERN experiment already implies a bound $ \Lambda
\gesim 10~\tev $.

If the previously mentioned factor of $ m_\mu / v $ is included in $ \alpha_{
\mu W , B } $, the above estimate decreases to $ \anewd \sim 4 \times 10^{-10
}/\Lambda^2 $. The sensitivity is accordingly diminished to $ \Lambda \lesim
1~\tev $ for the Brookhaven experiment, while the CERN data implies only that
$\Lambda \gesim 0.2~\tev $.

The suggestion here is that, if $\Lambda$ is of the order of the weak scale,
these effects may be as large as the Standard Model electroweak contributions.
SUSY models\foot{These are summarized by Kinoshita and Marciano in
Ref.~\km.} exemplify that possibility. Nevertheless, these limits, as we
mentioned previously, must be cautiously interpreted, since, in other models,
small coupling constants or resonances in the underlying theory can alter these
limits on $\Lambda$ by an order or magnitude or more. It is important to note,
however, that even if an effect is seen in the Brookhaven experiment, it will
not be produced by any modification of the ``anomalous" three-gauge-boson
couplings: $ \anewi + \anewii $ would produce a measurable effect only for
scales of a few GeV, corresponding to a region already probed by LEP, Fermilab
and many other previous experiments \REF\pdg{{\sl Particle Data Group,
Review of Particle
Properties}, {\sl Phys. Rev.} {\bf D45} (1992) sect. V.} [\pdg], which found
no evidence of new physics.

\chapter{{\caps Non-decoupling case}}

We now turn our attention to the possibility that physics above the scale
$\Lambda$ does not decouple from the low-energy physics. In this case the
appropriate expansion of the effective lagrangian is in powers of momentum. We
will assume here that the particle spectrum is the same as the Standard Model's
with the exception of the Higgs, so that the effective lagrangian can be
written as a gauged chiral model \REF\ab{T. Appelquist and C. Bernard, {\sl
Phys. Rev.} {\bf D23} (1981) 425.  H. Georgi, Ref. \georgi.} \REF\long{A.
Longhitano, {\sl Nucl. Phys.} {\bf B188} (1981) 118.} [\gl, \ab, \long].
In this model we expect that $\Lambda \sim 4\pi v $ [\ab], and that, for
energies small compared to $\Lambda$, the first terms in the expansion will
provide a good approximation [\georgi].

Specifically, if $$ U=\exp \left[ 2i \pi^a \tau^a / v \right] \eqn\eq $$ then
lowest order kinetic terms in the effective lagrangian are [\holdt] $$ \lcal
\lowti{ kin } = { v^2 \over 4 } \tr \{ \DD_\mu U^\dagger \DD_\mu U \} - \half
\tr \{ \WW_{ \mu \nu } \WW^{ \mu \nu } \} - \half \tr \{ \BB_{ \mu \nu } \BB^{
\mu \nu } \} \eqn\lkin $$ where we have adopted the matrix notation $ \WW_\mu =
W^I_\mu \tau^I / 2 $, $\BB_\mu = B_\mu \tau_3 / 2 $, $ \DD_\mu U = \partial_\mu
U + i g \WW_\mu U - i g' U \BB_\mu$, and $ \tau^I $ denote the Pauli matrices.

There are six new $ \su 2_L \times \ui_Y $ operators which are of chiral
dimension four or lower and contain quadratic or trilinear gauge vertices
[\long]. The only term of chiral dimension two is $$ \lcal_1'= {v^2 \over 4}
\beta_1' \left( \tr \left[ \tau^3 U^\dagger \DD_{\mu} U \right] \right)^2
\eqn\eq $$ and the five of order [mass]$^4$ are $$ \eqalign { \lcal_1 &= g g'
\beta_1 \tr \left[ U \BB_{ \mu \nu } U ^\dagger \WW^{ \mu \nu } \right] \cr
\lcal_2 &= - 2 i g' \beta_2 \tr \left[ \BB_{ \mu \nu } \DD^\mu U ^\dagger
\DD^\nu U \right] \cr \lcal_3 &= - 2 i g \beta_3 \tr \left[ \WW_{ \mu \nu }
\DD^\mu U \DD^\nu U^\dagger \right] \cr \lcal_8 &= \quarter g^2 \beta_8 \left(
\tr \left[ U \tau^3 U^\dagger \WW_{ \mu \nu } \right] \right)^2 \cr \lcal_9 &=
- i g \beta_9 \tr \left[ U \tau^3 U^\dagger \WW_{ \mu \nu } \right] \tr \left[
\tau^3 \DD^\mu U^\dagger \DD^\nu U \right]. \cr} \eqn\chirall $$ The numbering
system, prefactors and signs are adapted from those of Ref. \long. $\beta_1$
corresponds to $L_{10}$ of Ref. \gl, $\beta_2$ and $\beta_3$ to $L_9$ of that
same reference. $\beta_1'$ is denoted $\Delta_\rho$ in [\holdt].

As in the decoupling case, except for $\lcal_1',$ the naive order of
magnitude estimate of each $\beta$ is $ v^2/\Lambda^2 \simeq 1/ 16 \pi^2 $
[\georgi].  The $\beta$'s may be larger than expected, for example in
technicolor theories, where they are enhanced by the numbers of generations
and technicolors [\holdt], or if enhanced by a low-lying resonance.
$\beta_1'$, though {\it a priori} of order 1, violates the approximate
$SU(2)_R$, and can be limited by measurements of $\rho =  (M_W / M_Z
cos\theta_w)^2$ to $\beta_1' \lesim 1\%$\REF\plml{P. Langacker and M. Luo,
Phys.\ Rev.\ {\bf D44} 817 (1991.)} [\plml], which, coincidentally, is of the
same magnitude as $1/16\pi^2$.\foot{Once the top mass is known, the error on
$\rho$ will be reduced to a few tenths of a per cent.}

Just as in the last section we must also include the operators \directo\ which
give a direct tree-level correction to $\amu$, namely
$$
\lcal \lowti{ direct}
= {1\over v}\bar\psi_R\sigma_{\mu\nu}{\bf m} \psi_L
\left[ g\beta_{\mu W} W^{\mu\nu} + g' \beta_{\mu B} B^{\mu\nu} \right],
\ \ \ \ {\bf m} =  \pmatrix{0&0\cr 0&1\cr}.
\eqn\dirnon
$$
The $\beta$'s in the direct terms are expected to be of order $1 / 4\pi$. We
will consider separately the cases where $ \beta_{\mu W, B } $ are decreased by
a factor $ m_\mu / \Lambda $. Our complete effective lagrangian is then given
by expressions \lkin\ through \dirnon.

We look first at the direct terms. A quick calculation shows them to be $$
\anewd = { g m_\mu \over \sqrt{ 2} M_W } ( \beta_{ \mu W } - \beta_{ \mu B } )
\eqn\directnd$$

As in the decoupling case, there will be contributions to $ \anew $ from
three-boson vertices and from two-boson vertices; $\lcal_1'$, $\lcal_1$, and
$\lcal_8$ have bilinear terms, and all but $\lcal_1'$ have trilinear terms. We
first consider the trilinear terms. To this order in the expansion of $ \lef $
there is no contribution like $\alpha_W$ since $ \ocal_W $ has chiral dimension
six. We can see by comparison that the contributions from $\beta_1$, $\beta_2$,
$\beta_3$, $\beta_8$, and $\beta_9$ are all proportional to the $\alpha_{ W B
}$ term of the decoupling lagrangian so that we need only make the replacement
$$ \alpha_{ W B } { M_W^2 \over \Lambda^2 } \rightarrow \quarter g^3 g' \left[
-2 \beta_1 + \beta_2 + \beta_3 - \beta_8 + \beta _9 \right] \eqn\eq $$ and,
therefore (using $ \overline{ M S } $ subtraction and taking $v$ as the
renormalization scale) $$ \anewi = { g^4 \over 8 \pi^2 } { m_\mu^2 \over M_W^2
} \left[ -2 \beta_1 + \beta_2 + \beta_3 - \beta_8 + \beta _9 \right] \left( \ln
{v^2 \over M_W^2 } + {3 \over 2} \right) . \eqn\ind $$ The ultraviolet
divergence is, as in the decoupling case, non-observable and has been cancelled
by the appropriate counterterms in $ \lcal_ { \mu W } + \lcal_{ \mu B }$.

Next we must consider the effect of the terms quadratic in gauge bosons on $
\anew $. The quadratic part of the gauge-boson lagrangian is $$ \eqalign {
\lcal^2_{ W , B }= & -\quarter W_{ \mu \nu }^I W^{ I \mu \nu } + {g^2 \over 4}
\beta_8 W_{ \mu \nu }^3 W^{ 3 \mu \nu } - \quarter B_{ \mu \nu } B^{ \mu \nu }
+ \half g g' \beta_1 W_{ \mu \nu }^3 B^{ \mu \nu } \cr & + { v^2 \over 8} ( g^2
W_\mu^I W^{ I \mu }+ g'{}^2 B_\mu B^\mu - 2 g g' W_\mu^3 B^\mu ) \cr & -
\beta_1' {v^2 \over 4} ( g W_\mu^3 - g' B_\mu ) ( g W^{ 3 \mu } - g' B^\mu ).
\cr } \eqn\eq $$ This requires the re-diagonalization $$ \eqalign { W^3_\mu & =
\sw \left[1 + \half g^2 \sw^2 (2 \beta_1 + \beta_8) \right]A_\mu + \cw \left[ 1
- {g'}^2 \sw^2 \beta_1 + \half g^2 ( 1 +\sw^2) \beta_8 \right] Z_\mu \cr B_\mu
& = \cw \left[ 1 + \half g^2 \sw^2 (2 \beta_1 + \beta_8) \right] A_\mu - \sw
\left[ 1 - \half g^2 \cw^2 (2 \beta_1 + \beta_8 ) \right] Z_\mu . \cr } \eqn\eq
$$ These expressions lead to the redefinitions $$ \eqalign{ e \rightarrow e^*
=& e \left[ 1 + \half g^2 \sw^2 (2 \beta_1 + \beta_8 ) \right] , \cr M_Z
\rightarrow M_Z^* =& M_Z \left[ 1 - \beta_1' - \half g^2 ( 2 \sw^2 \beta_1 -
\cw^2 \beta_8 ) \right] ; \cr } \eqn\eq $$ while $ G_F $ and $ M_W $ remain
unchanged.

As in the previous case the above expressions produce a new term in the
Standard Model contribution to $ \amu $ due to the modification in $M_Z$ and
the $ Z \bar \mu \mu $ couplings, which in turn modify $ \amu$. A
straightforward calculation gives $$ \anewii = { G_F m_\mu^2 \over 3 \sqrt{2}
\pi^2 } \left[ (1- 4 \cw^4 ) \beta_1'- g^2 ( 1 - 4 \sw^2) \left( \sw^2 \beta_1
- \cw^2 \beta_8 \right) \right]. \eqn\iind $$

The total change in $\amu$ is again given by the sum $ \anewd + \anewi +
\anewii $, where these quantities are given in \directnd, \ind\ and \iind\
respectively. Numerically $ | \anewi + \anewii | \lesim 6 \times 10^{-10 } $
depending on the relative signs of the $ \beta_i $. The direct contribution is
restricted to $ | \anewd | \lesim 10^{-4} $ if no factor of $ m_\mu /
\Lambda $ appears in $ \beta_{ \mu B , W } $, or $ | \anewd | \lesim 3 \times
10^{-9}$ if this factor is present.

We see then that the situation here is marginally different from the decoupling
case: if the $ \beta_i $ all conspire to suppress the direct contribution and
enhance (by a factor of two or so) $ \anewi + \anewii $, then the Brookhaven
experiment may be sensitive to the effects of an anomalous triple-gauge-boson
vertex. On the other hand, if we assume no significant cancellations between
the $ \beta_i$ and no unexpected enhancement of these coefficients, then the
main contribution to $ \anew $ comes from $ \anewd $, just as in the decoupling
case, while the other contributions are unobservable. In this case, current
CERN data implies a very strong suppression of the direct contributions which
can be  interpreted naturally as evidence for the factor $ m_\mu / \Lambda $
in $ \beta_{ \mu B , W }$; when this is included the contribution from these
terms lies below the sensitivity of the existing data (but well inside that of
the Brookhaven experiment). We can conclude that, if the non-decoupling case is
realized in nature, the ``direct" contributions must be suppressed by a $m_\mu/
\Lambda$ factor and that the Brookhaven experiment will either observe them or
set interesting limits, implying that these contributions are further
diminished (for example, by arising only at two-loops).

The discussion above includes the $\alpha_{WB}$ (or, in the conventional
notation\foot{See next section for the notational relations.}$\kappa-1$,) piece
of the triple-gauge-boson vertex. The $\alpha_W$ (or $\lambda$, in the
conventional notation) part is unchanged from the decoupling case and is
therefore probably unobservable.

\chapter{{\caps Comparison to other results}}

Several authors [\km, \long] have considered the effects of high-energy
physics on the anomalous moments of the W. In this section we compare these
results with ours in the decoupling scenario.

In our notation the effects considered in [\km, \long] are described by
the replacements $$ \alpha_{ W B } = ( \kappa -1 ) g g' \Lambda^2 / ( 4 M_W^2 )
\qquad \hbox{and} \qquad \alpha_W = \lambda g \Lambda^2 / ( 6 M_W^2 ); \eqn\eq
$$ all other $\alpha_\ocal $ are ignored. The constants $\lambda$ and $\kappa$
are related to the magnetic dipole and electric quadrupole moments of the W by
$$ \mu_W = { e \over 2 M_W }( 1+ \kappa + \lambda ) \qquad \hbox{and} \qquad
Q_W = - {e \over M_W^2 }( \kappa - \lambda ) . \eqn\eq $$ Note that our
previous arguments imply that, taking the most benign case where $ \Lambda = v
$, the natural scale for these constants is $| \kappa -1 | \sim 3 \times 10^{ -
3 } $ and $ | \lambda | \sim 2 \times 10^{-3} $. With this proviso, the results
obtained in [\km, \long] coincide with $ \anewi $ in \amudec.

\chapter{{\caps Conclusions}}

We have described the formalism of loop calculations for effective-lagrangian
models, using $ \amu $ as an example. The philosophy of our approach differs
markedly from the one used in several other publications [\km], and this
translates into different conclusions. Firstly, we note that any divergence
obtained in using an effective lagrangian is unobservable since it will always
be cancelled by appropriate counterterms appearing in other operators in $ \lef
$. The only remnant of these divergences concerns the logarithmic ones which
specify the renormalization group flow of the couplings due to the light
fields. It is only in this sense that the logarithms in \amudec\ are
observable. Stronger divergences\ \REF\kvy{G.L. Kane, J. Vidal, and C.P. Yuan,
{\sl Phys. Rev.} {\bf D39} (1989) 217. G. Altarelli and R. Barbieri, {\sl
Phys. Lett.} {\bf B253} (1991) 161. A. Frau and J.A. Grifols, {\sl Phys.
Lett.} {\bf B166} (1986) 233. R. Alcorta, J.A. Grifols, and S. Peris, {\sl
Mod. Phys. Lett.} {\bf A2} (1987) 23. J.J. van der Bij, {\sl Phys. Rev.} {\bf
D35} (1987) 1088.} \ (quadratic or quartic) are completely unobservable; this
argument contrasts with several other opinions [\kvy]. Related to this issue
is the constraint of gauge invariance; this allowed us to use a
gauge-preserving regularization where divergences higher than logarithmic are
automatically (and consistently!) disregarded.

Secondly, it is important to note that the magnitudes of the dimensionful
coefficients reflect assumptions regarding the scale at which new physics will
become apparent. For example if we require $ \lambda \sim 1 $, this implies a
scale of $ \sim 10\ \gev$ which is obviously irrelevant; similar results hold
if $ | \kappa - 1 | \sim 1 $. It is important to note that the scale $ \Lambda
$ in the logarithms and the one multiplying the prefactors must be the same
(required by consistency) and the observability limits cannot ignore this fact.

These conclusions apply independently of the nature of the physics beyond the
Standard Model, whether confining or weak. For example, in the approach
advocated in this paper, the statement that $ \lambda = O ( 1 ) $ in a
composite model is untenable.

Finally we remark on the estimates we used for the couplings. In the decoupling
case we have assumed that each gauge boson is associated with a coupling
constant $g$ or $g'$ and that, since all the operators have dimension larger
than 4, they represent the low energy limit of a series of loop diagrams; this
relies to a certain extent on perturbation theory. But in the case where the
underlying physics does not lie in the perturbative regime we can borrow the
arguments used in the chiral approach to the strong interactions [\georgi],
which lead to similar results. The non-decoupling case again closely
parallels QCD, and we use the corresponding arguments. If the factors of $ 1 /
16 \pi^2 $ in our estimates are ignored, the magnitude of the contributions
increases about two orders of magnitude and the conclusions are markedly
altered. Since we have no reason to suppose such an anomalously large value
for the $ \alpha_\ocal $ or the $ \beta_i $ we have not considered this
possibility.

Effective lagrangians can be used to calculate observables in a loop expansion.
Among other things, these loop contributions modify the direct terms,
contributing to the $\beta$-functions for $\alpha_{ \mu W }$ and $\alpha_{ \mu
B }.$ Whether these ``direct" terms are larger than suggested by operator
mixing is model dependent, as we discussed followed Eq.~\estimates. Taken at
face value, in the decoupling case, the BNL experiment will push the limits on
$\Lambda$ from their present value of $200~GeV$ up to about $1~TeV.$ However,
it may well be that the relevant threshold associated with new physics is much
lower than $\Lambda,$ depending on the nature of the underlying theory. In the
non-decoupling case, the Brookhaven experiment should again be sensitive to the
underlying physics whose effects should be apparent at scales of order $v$. In
the decoupling case, certain models ([\vol], SUSY) illustrate the possibility
that
the direct couplings are not suppressed by a $ m_\mu / \Lambda $ factor (an
unlikely possibility for the non-decoupling scenario in view of the CERN data)
in which case the sensitivity of the Brookhaven experiment is increased to
$\Lambda \sim 50~\tev $. For both situations however, the measurements of $
\amu $ will not be sensitive (or at best marginally so) to anomalous
triple-gauge-boson vertices.

\REF\bl{C.P. Burgess and D. London, McGill University reports 92/04 and 92/05
(unpublished)} After this manuscript was completed we became aware of Ref.
\bl, in which many of the points of principle discussed above are also
addressed.

\ack

Portions of this work were supported by the
Department of Energy, and by the
National Science Foundation (Grant No. PHY89-0435).

\endpage
\refout
\endpage
\bye